\documentclass[a4paper,10pt]{article}
\usepackage{graphicx}

\title{Using diffuse radio emission in clusters of galaxies to probe the cosmic web, cosmic rays and dark matter\footnote{Proceedings of the ICRC (The Astroparticle Physics Conference) 2013, The Brazilian Journal of Physics, in press}}

\author{Nonis$^{1}$, Stavros and Gizani$^1$, Nectaria A. B., for the ICRC Collaboration\\
$^1$ Physics Laboratory, School of Science and Technology, Hellenic Open University, \\ Patra, Greece \\}

\begin{document}
\maketitle

\paragraph{Abstract}

The study of diffuse radio sources is essential to our knowledge of the physical conditions in clusters of galaxies and of the role that large scale magnetic fields play in the propagation of the relativistic particles in the intracluster medium (ICM). 
Due to dissipative processes, the thermal plasma of the ICM contains useless information about the formation of structures and CRs. Such past and current information is carried intact by its non-thermal emission caused by highly relativistic particles spiraling in large-scale magnetic fields. The diffuse radio emission is not associated with the active galactic nucleus phenomenon and has no optical counterpart. Instead it is associated with the ICM and it is characterized by steep spectrum. The low surface radio brightness features are thought to be formed via reacceleration of a relic population of relativistic electrons, or proton-proton collisions with the ICM. Diffuse radio sources are classified typically as `relics' , `halos', 'mini-halos' and 'radio-ghosts'. 
We use a sample of 67 nearby clusters of galaxies at redshift, 0.02 $<$ z $<$ 0.1, to tackle the relation between the diffuse radio emission and the large scale structure (cosmic web), dark matter (DM) and cosmic rays (CRs). 
Our sample of clusters at low redshift containing areas of radio flux below the Jy level is well suited for the study of DM. DM halos can substantially contribute to the counts of sources at low brightness.
We report on the work in progress.\\


\section{Introduction}

Clusters of galaxies are the most massive gravitationally bound systems in the Universe. They are also the nodes in the large scale filamentary cosmic structure formed by hierarchical growth. Typical cluster radii are of a few Mpc and total masses of $\sim (10^{14}-10^{15})$M$_{\odot}$ \cite{bib:Voit}.
Most of the gravitating matter in any cluster is in the form of dark matter ($\sim$ 80\%). Some of the luminous matter is in galaxies ($\sim$ 3-5\%), the rest is in diffuse hot gas ($\sim$ 15-17\%)  \cite{bib:Voit}. The hot gas, or ICM, has been detected in galaxy cluster cores through its thermal bremsstrahlung X-ray emission. Its typical temperatures are (1-10) keV, electron densities  $(10^{-2} - 10^{-3})$ cm$^{-3}$ \cite{bib:Ferrari}, and X-ray luminosities can be up to L$_x\sim 10^{45}$ erg s$^{-1}$ (see \cite{bib:Ferrari} for a review).

In the hierarchical scenario of the formation of cosmic structure, galaxy clusters are believed to form through anisotropic and episodic accretion of mass and the number of irregular and merging clusters increases at high redshift. Indeed a large fraction of clusters shows evidence of substructures both in their galaxy distribution (optical data) and in the X-ray emission morphology (through the ICM). Direct evidence of merger events was first provided by the ROSAT \cite{bib:43} and ASCA \cite{bib:44} satellites, suggesting that clusters are still forming at the present epoch as expected from the hierarchical formation model. The high angular and spectral resolution of the Xray telescopes (including Chandra and XMM) allow resolving the internal structure of clusters, and identifying signatures of merger activity: surface brightness discontinuities associated with jumps (shocks and fronts) in gas density and temperature. Such energetic phenomena may also show up in the non-thermal, diffuse radio emission of clusters of galaxies.

Observationally X-ray clusters can be divided into two classes: cool cores (CC, cooling flow clusters) with dense gaseous core regions in which gas temperature drops inwards, and non-cool cores (NCC) with shallower core profiles \cite{bib:40}. CC clusters have sharply peaked central X-ray emission from the hot dense gas which allows efficient central cooling. These clusters also have relatively undisturbed ICM, indicating that they are fairly dynamically relaxed. On the other hand, NCC clusters typically show more substructures. Present observations support an evolutionary scenario \cite{bib:41}, in which galaxy clusters naturally evolve into a CC state, until a merger event occurs. Then the energy released by the merger can efficiently disrupt the cool core, leading to NCC clusters which should be in an un-relaxed, post-merger stage \cite{bib:42}.

Radio relics, halos, mini-halos and radio-ghosts are non-thermal, diffuse, low surface brightness, extended radio sources with steep spectrum and no apparent cut off. They have no optical counterpart. They are not associated with the active galactic central engine, but with the ICM. Observational results provide evidence that these phenomena are related to cluster merger activity, which supplies energy to reaccelerate the radiating particles in the ICM. The halo-merger connection is investigated by comparing the radio and X-ray emission from these structures. 

\begin{figure}[pb]
\begin{center}
\includegraphics*[width=10cm]{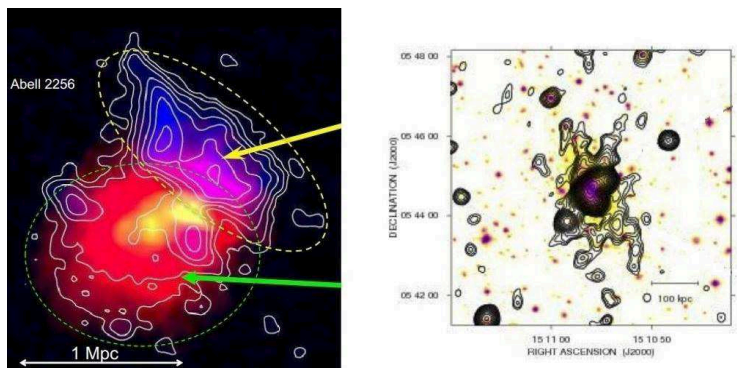}
\end{center}
\caption{Left: A VLA radio halo with size $\sim$1 Mpc (green arrow) at 1.4 GHz (contours) overlaid on the X-ray Chandra image of the host cluster A2256 from our sample. The $\sim$ Mpc size relic with elongated filamentary morphology at the cluster's outskirts is shown by the yellow arrow (image from \cite{bib:47}, with for H$_{\circ}$=71 km s$^{-1}$ Mpc$^{-1}$, q$_{\circ}=0.5$). Right: Total intensity radio contours of mini-halo in our sample cluster A2029 at 1.4 GHz overlaid on the Chandra X-ray image in the 0.5-4 keV band (map from \cite{bib:23}.}
\label{pic1}
\end{figure}

The classification of these radio sources is based on their location in the cluster and the cluster type: Structures at the cluster centre are called halos, at the  periphery of both merging and relaxed clusters are named relics or "radio gischt", while at the centre (also at intermediate distance) of relaxed cool core clusters are the mini-halos \cite{bib:Feretti}. Radio halos (Figure \ref{pic1}, left) have $\sim$ Mpc sizes (for H$_{\circ}$=71 km s$^{-1}$ Mpc$^{-1}$, q$_{\circ}=0.5$), display smooth morphologies and are usually unpolarized \cite{bib:Feretti}. Radio relics (Figure \ref{pic1}, left) are usually elongated and are suggested to be related to the large scale structure, i.e. to the filaments where galaxy clusters lie. Their radio emission is polarized with inhomogeneous and aspherical spatial distribution. Mini-halos (Figure \ref{pic1}, right) are extended on scales of $\sim$ 500 kpc (for H$_{\circ}$=71 km s$^{-1}$ Mpc$^{-1}$, q$_{\circ}=0.5$) and their sizes are comparable to the cooling region size \cite{bib:Feretti},  they have low surface brightness and present steep spectrum like halos and relics. Almost always mini-halos are found at the cluster centre, where powerful radio galaxies are hosts. Their radio detection requires high resolution  and dynamic range observations, in order to distinguish between their diffuse emission and the central radio galaxy's emission.   

Although the basic observational properties of radio halos are known their formation mechanism is still under study \cite{bib:Feretti}. Two dominant scenarios for their formation exist. The first one is the so-called "primary" model \cite{bib:Brunetti} in which an existing electron population is reaccelerated by turbulence or shocks caused by recent cluster mergers. The other is the "secondary" model in which relativistic electrons are continuously injected into the ICM by inelastic collisions between cosmic rays and thermal ions \cite{bib:Keshet}. Combinations of both acceleration mechanisms have also been considered \cite{bib:45}. At present the favored scenario for the formation of radio relics that they are the signatures of (re-)accelerated electrons at shocks due to cluster mergers \cite{bib:van Weeren}. The origin of mini-halos is poorly understood. Gitti et al.\cite{bib:Gitti} argued that radio emitting particles in mini-halos cannot be connected to the central radio galaxy in terms of particle diffusion. They proposed that mini-halos result from a relic population of relativistic electrons reaccelerated by turbulence via Fermi-like processes with the necessary energetics being supplied by the cool-core region.

Various astrophysical observations suggest that most of the Universe matter content is in the form of cold DM. However the precise nature of DM is one of the most important and still open questions in modern physics. Many different candidates have been proposed as DM constituents, but there is no evidence in favour of any model. One of the most studied scenarios is that of weakly interacting massive particles (WIMPs) for which the most representative candidate is the supersymmetric lightest particle, namely the neutralino (see \cite{bib:Bertone} for a review).

In general, WIMPs can self-annihilate or decay. In the context of annihilating DM, WIMPs are favored by the fact that they naturally have a relic density that matches the observed DM abundance. For decaying DM, it has been shown that WIMPs can have decay lifetimes larger than the age of the Universe, and are therefore viable DM candidates. The secondary products of the DM decay or annihilation can generate non-thermal emission from radio to gamma-ray frequencies (through various mechanisms like synchrotron emission, Inverse Compton Scattering etc). Therefore another possibilty for the origin of the relativistic particles (electrons and positrons) responsible for the diffuse radio emission could be the products of dark matter annihilation.  

In our research work and throughout this paper we assume a $\Lambda$CDM cosmology with the latest Planck results, H$_\circ$=67 km s$^{-1}$ Mpc$^{-1}$, $\Omega_{m}=0.32$ and $\Omega_{\Lambda}=0.68$ \cite{bib:46}.

\section{The Sample}

Initially our sample consisted of 67 Abell clusters of galaxies at redshift 0.02 $<$ z $<$ 0.1. The clusters were selected based on their X-ray and optical/near-IR properties. First we have searched for diffuse radio emission (in the form of either radio halo/relic or mini-halo) among the clusters using the literature (see Table \ref{table_1}). 13 clusters were found to present diffuse radio emission (radio halos/relics/mini halos). They have been studied extensively both in the X-ray and radio band, searching for merger activity and diffuse radio emission respectively. 12 clusters are relaxed (i.e. no evidence for merger activity), while 3 of them (A1413, A1650, A2029) host mini-halos. 27 clusters have not been observed either in radio or X-rays, and we are going to use the VLA to gather radio data. 

Table \ref{table_1} presents the sample we are left with, after excluding the clusters that are confirmed to be relaxed. Clusters that their X-ray observations show evidence of merger activity (e.g. substructure in the gas density distribution, multiple X-ray peaks, distortions, edges and tails of X-ray emission, patchy temperature structures, metallicity gradients) are displayed under the title 'MA' denoting 'Merger activity', while clusters containing diffuse radio sources are listed under 'DRE' implying 'Diffuse Radio emission'. The diffuse structure type (RH: radio halo, RR: radio relic, MH: mini halo) is also included in the same column. The observing radio frequency 'f' is presented in the last column.

\begin{table}[h]
\begin{center}
\begin{tabular}{|c|c|c|}
\hline MA & DRE, Type & f (MHz) \\ \hline
A85 \cite{bib:1,bib:2} & RH & 300 \\ \hline
A133 \cite{bib:4,bib:5}& RR& 1400 \\ \hline
A154 \cite{bib:6}&  - & \\ \hline
A168 \cite{bib:20} & -  & \\ \hline
A399 \cite{bib:7,bib:8}& RH & 1400 \\ \hline
A401 \cite{bib:7,bib:8}& RH & 1400 \\ \hline
A500 \cite{bib:9}& - & \\ \hline
A514 \cite{bib:10}& - & \\ \hline
A665 \cite{bib:11,bib:1} & RH &  1400 \\ \hline
A957 \cite{bib:49}& - & \\ \hline
A1213\cite{bib:1} & RH & 1400 \\ \hline
 & A1413 \cite{bib:13}, MH & 1400 \\ \hline
A1569 \cite{bib:15} & - & \\ \hline
 & A1650 \cite{bib:12,bib:16}, MH & 1400  \\ \hline
A1656 \cite{bib:17,bib:18}& RH, RR & 1400  \\ \hline
A1775 \cite{bib:19}& - &   \\ \hline
A1913 \cite{bib:21} & -  & \\ \hline
 & A2029\cite{bib:22,bib:23}, MH & 1400 \\ \hline
A2255 \cite{bib:26} & RH, RR & 325   \\ \hline
A2256 \cite{bib:27,bib:28,bib:Bonamente}& RH, RR & 327   \\ \hline
A2384 \cite{bib:30}& - &   \\ \hline
A2440 \cite{bib:30}& -  & \\ \hline
A2626 \cite{bib:34,bib:35}& MH & 1500  \\ \hline
A2670 \cite{bib:37}&- &  \\ \hline
A2804 \cite{bib:9}&- &  \\ \hline
A2933 \cite{bib:30}& -  & \\ \hline
A3128 \cite{bib:39} &- &\\ \hline
A3158 \cite{bib:34,bib:35}&- & \\ \hline\
A3223 \cite{bib:9}&- & \\ \hline
A3266 \cite{bib:51}&- & \\ \hline
A3921 \cite{bib:52}&- &\\ \hline
\end{tabular}
\caption{The clusters of our sample that present diffuse radio emission and/or merger events. For details, see text.}
\label{table_1}
\end{center}
\end{table}

\section{The Methodology}

We have gathered radio VLA, VLBA, and ROSAT-CHANDRA X-ray data for the clusters that are confirmed to have a merger activity. We collect gamma ray data from INTEGRAL, VERITAS and possibly in the future from FERMI satellite. Optical and Infrared data will also be gathered for the sources of interest. 

Our scope is to find correlations between the cluster gamma-, X-ray luminosity and the radio power of the halos, relics and mini-halos. 

\section{Discusssion and Future Work}

Diffuse radio sources imply non-thermal radiative processes caused by highly relativistic particles spiraling in large-scale magnetic fields. Therefore they  are indicative of the presence of external magnetic fields and particles in the ICM as well as of the physical conditions present in clusters of galaxies.   

Radio relics/halos and similar features are thought to be formed via reacceleration of a relic population of relativistic electrons, or proton-proton collisions with the ICM.  Radio halos are also related to merger events. They are found in the cluster center.  Most clusters of galaxies with cool cores have active galactic nuclei at their centres. These AGN can produce bubbles of non-thermal radio-emitting particles (lobes). One explanation for the formation of `radio ghosts'  detected far from the cluster centres could be the magnetic confinement of cosmic rays by these bubbles.

The aim of this research is to use magnetic field information and correlate  the cluster gamma-ray, X-ray luminosity and radio power of the halos/relics and radio ghosts in order to examine: (a) Models suggesting a hadronic origin of the radio halos. Cosmic ray protons generate -rays, neutrinos, and secondary electrons and positrons through pion-collisions with gas atoms. Any synchrotron or inverse Compton radiation from secondary electrons and positrons would therefore be  accompanied by -rays induced by pion-decay. Gamma-rays may also be produced in acceleration sites of ultra-high energy cosmic rays, as a result of interactions with the cosmic microwave background (CMB), (b) whether the presence of radio ghosts indicates suppression of the CR diffusion and  their partial escape. This effect would produce observable gamma-rays via the interaction of the diffusing CRs with the thermal ICM, (c) We constrain the properties of the clusters containing non-thermal radio emission, (d) the shocks produced in the intergalactic medium during large-scale structure formation. They are thought to accelerate a population of highly relativistic electrons by the intergalactic magnetic fields and therefore they emit in the radio regime. (e) Cast list in the distributions of Galactic CRs, ionized gas and magnetic fields.

We adress DM in a non-direct way. Residual self-annihilation in the DM can give rise to significant fluxes of $\gamma$-rays, electrons, positrons, and neutrinos, which can be detected indirectly. In addition dense regions of dark matter can produce some antimatter such as anti-protons and positrons.  Secondary electrons and positrons can in turn annihilate and be accelerated in the magnetic fields of galaxies and can be detected in the radio band. 

Previous works suggest that extragalactic Weakly Interacting Massive Particles (WIMPs), predicted by supersymmetric extensions of the Standard Model (e.g. neutralinos), emit radio emission which decreases with redshift. That is the main contributions come from structures at z $\ll$ 1. The population should also peak at low radio brightness. In addition extragalactic radio background from normal and radio galaxies dominate down to kHz frequencies, while CMB  dominates the radio sky at frequencies above $\simeq$ 1 GHz. Our sample of clusters at low redshift containing areas of radio flux below the Jy level is well suited for the study of DM. We look for  DM halos since they can substantially contribute to the counts of sources at low brightness. In a later stage we will be able to test this hypothesis with LOFAR observations at low frequencies.  
Future gamma ray observations of the clusters of interest would also be useful to our research work. 

\vspace*{0.5cm}
\paragraph{Acknowledgments}
NG wishes to acknowledge ICRC 2013 for the financial support she has been granted

\end{document}